\documentclass[sigconf]{acmart}

\usepackage{graphicx}
\usepackage[utf8]{inputenc}  
\usepackage{graphicx,url}
\usepackage{colortbl}

\usepackage[]{algorithm2e}

\usepackage{booktabs}
\usepackage{siunitx}
\usepackage[flushleft]{threeparttable}
\usepackage{multirow}
\usepackage{etoolbox}
\usepackage{makecell}
\usepackage{array, tabularx, boldline}
\usepackage{cellspace}
\usepackage{algorithmic}

\setcopyright{none}

\newcommand{\ROGUEStatement}{social relationship mining process}

\newcommand{\RECAST}{RECAST}

\newtoggle{HIDDEN}
\toggletrue{HIDDEN}
\definecolor{fixclr}{rgb}{0.0, 0.5, 0.0}

\newcommand{\iAPS}{\textit{APS}}
\newcommand{\iDBLP}{\textit{DBLP}}
\newcommand{\iPUBMED}{\textit{PubMed}}
\newcommand{\iDARTMOUTH}{\textit{Dartmouth}}
\newcommand{\iUSC}{\textit{USC}}
\newcommand{\iENRON}{\textit{Enron}}

\begin{document}
\title{Improving Community Detection by Mining Social Interactions}

\author{Jeancarlo C. Le\~{a}o}
\orcid{0002-8181-7965}
\affiliation{%
  \institution{Instituto Federal do Norte de Minas Gerais}
  \city{Araçuaí}
  \state{Minas Gerais}
  \country{Brazil}
  \postcode{31840-030}
}
\email{jeancarlo.leao@ifnmg.edu.br}

\author{Michele A. Brand\~{a}o, Pedro O. S. Vaz de Melo, Alberto H. F. Laender}
\orcid{0002-8181-7965}
\affiliation{%
  \institution{Department of Computer Science \\ Universidade Federal de Minas Gerais}
  \streetaddress{Av. Antônio Carlos, 6627, Pampulha}
  \city{Belo Horizonte}
  \state{Minas Gerais}
  \country{Brazil}
  \postcode{31840-030}
}
\email{{micheleabrandao,olmo,laender}@dcc.ufmg.br}

\renewcommand{\shortauthors}{Leão. Jeancarlo et al.}

\begin{abstract}
Social relationships can be divided into different classes based on the regularity with which they occur and the similarity among them. {Thus, rare and somewhat similar relationships are random and cause noise in a social network, thus hiding the actual structure of the network and preventing an accurate analysis of it.} In this context, in this paper we propose a process to handle social network data that exploits temporal features to improve the detection of communities by existing algorithms. By removing random interactions, we observe that social networks converge to a topology with more purely social relationships and more modular communities.
\end{abstract}



\maketitle

\section{Introduction}

Many studies on community detection in temporal social networks use aggregated static graphs due to the difficulty of considering temporal aspects \cite{greene2010tracking}. However, 
this simplification causes informational noise in the social relations, which can lead to errors in the individuals' membership in their respective communities. For example, consider a group of people who do not know each other and exchange many emails in a single day, but after that they do not communicate again. Now consider another group of people who exchange many messages regularly over years. 
Although the relationships between the members of both groups have the same topology when considering a static network, the temporal dimension allows one to differentiate the relationships and, consequently, the community structures 
involving these two groups of people.

In fact, networks formed by aggregating interactions within a dynamic system are subject to a wide variety of noise. This means that an edge or relationship may be randomly established between pairs of vertices representing individuals with a low probability of interaction. This is the case of emails sent to a wrong address or when a contact is just added due to a facility offered by a social media~\cite{Abufouda:2015:WRF:2740908.2742468}. However, randomness is also related to ephemeral relationships like, for example, casual contacts or coauthorships.
On the other hand, considering real communities, a fundamental property shared by different definitions\footnote{There is no universally accepted definition for the concept of community \cite{Fortunato201075,Palla2005}, which has proved to be difficult to define, quantify and extract \cite{Abrahao:2012:SSC:2339530.2339631}.} is the presence of real social relationships within a community, which are usually sustained over time. Thus, it is of paramount importance to consider the temporal dimension in the identification of true communities that are free from randomness and noise. Moreover, assessing how real a relationship is in a network is very important in order to get a high quality representation of its community structure \cite{Abufouda:2015:WRF:2740908.2742468}.

There are many approaches for detecting communities in  
networks~\cite{Abrahao:2012:SSC:2339530.2339631,blondel20081742-5468-2008-10-P10008,Clauset2004PhysRevE.70.066111,Raghavan2007,NewmanGirvanPhysRevE.69.026113,Xie2013}. Abrahao et al.~\cite{Abrahao:2012:SSC:2339530.2339631} and Xie et al.~\cite{Xie2013}, for instance, present a comprehensive analysis of the properties of communities detected by several algorithms. They show that the {detected communities} and their properties vary consistently across algorithms. In these two studies and in many others on community extraction, only the static {relationships are} analyzed.
Nevertheless, some community detection algorithms have been applied over temporal networks
~\cite{Liu201718449,1367-2630-11-3-033015,palla2007quantifying},
{which represent snapshots as a sequence of static graphs. In this case, the usual approaches detect communities in each snapshot independently \cite{palla2007quantifying} or iteratively}~\cite{1367-2630-11-3-033015}. Other algorithms consider the temporal aspect to identify dynamic communities by {globally detecting them in all snapshots~\cite{Liu201718449}.} Unfortunately, community detection approaches that exploit temporal aspects still comprise a very small part of current work when compared to those based on static networks.

Thus, instead of developing new algorithms for detecting communities, in this paper we propose a method to favor those that already exist, but that explore minimal properties. 
Specifically, we assess social relationships 
by measuring topological and temporal aspects on data extracted from a social network, which allows us to quantify the noise 
by using distinct algorithms.

In summary, in this paper we propose a \ROGUEStatement~that allows improving the quality of communities detected by existing algorithms. First, we show that our proposed method significantly increases the structural quality of the communities detected by several state-of-the-art algorithms.
Second, although there is no Holy Grail algorithm that solves the problem of community detection~\cite{Peele1602548}, our proposed method is able to increase the consensus among the results of several of these algorithms, i.e., the  communities detected by them become more similar. Finally, we show that our detected communities are more similar to their ground truths, when such information is available. 
For more details of the work presented in this paper, we refer the reader to its full version~\cite{Leao2018JISA}.

\section{Methodology}
\label{sec_methotodology}

The community detection problem can be summarized as follows. Given a non-directed graph ${G}{(V, E)}$, where $V = \{ v_{1}, ..., v_n\}$ is the set of vertices and ${E =}$ ${\{e_1, ..., e_m\}}$ is the set of edges representing the interactions between two vertices, the {community detection problem} consists of finding the set of non-overlapping communities $C = \{c_1, c_2, ..., c_k\}$ in which each vertex $v_i$ is associated with a community $c_i \in C$.
Considering a dynamic scenario, in which nodes interact with each other over time, it is possible to construct temporal graphs from interaction windows. More specifically, each temporal graph $G_t (V_t, E_t)$ in $G$ represents the aggregation of interactions in discrete periods of time $t$. Thus, for a given value of $t, V_t$ includes all vertices that interacted in the $t^{th}$ period of time. The edges in the set $E_t$ 
connect pairs of nodes ($v_i, v_j$) that interacted during the time period $t$.

In this paper, we apply this 
graph model to real social networks from three distinct domains: scientific collaboration networks established in the years 2000-2016 and built from the APS, DBLP and PubMed datasets\footnote{APS (\url{http://www.aps.org/}): collaboration network of the American Physical Society; PubMed (\url{https://www.ncbi.nlm.nih.gov/pubmed}): collaboration network from MEDLINE articles; DBLP (\url{http://dblp.org/}): collaboration network from DBLP computer science conferences.}, 
university campus mobility networks from Dartmouth College and USC~\cite{VazdeMelo2015}, and an email network derived from 
contacts among Enron's employees from 1999 to 2003\footnote{Enron email dataset: \url{https://www.cs.cmu.edu/~./enron}}. Table~\ref{table:networks_statistics} presents a 
brief characterization of these networks. In the scientific collaboration networks, nodes represent researchers and there is an edge between two researchers if they have coauthored a paper together. In the mobility networks, nodes correspond to members of a university community (e.g., students or faculty members) and there is an edge between two individuals 
if they have been both connected to a given Wi-Fi access point at the same time. Finally, in the email network, nodes are employees from Enron and there is an edge between them if they have exchanged emails. 

\setlength{\tabcolsep}{0.45em} 

\if 0
\begin{table}
        \sisetup{round-mode=places
        ,round-precision=0
        ,scientific-notation=true
    }
\centering
\begin{threeparttable}
    \caption{Characterization of the networks.}
    {
    \label{table:networks_statistics}
    \small
    \begin{tabular}{@{}lrrrrrrr@{}} 
    \toprule
\multicolumn{1}{c}{\emph{Network}} & \multicolumn{1}{c}{$V(k)$} & \multicolumn{1}{c}{$E$} & \multicolumn{1}{c}{$\Delta$} & \multicolumn{1}{c}{$\alpha$} & \multicolumn{1}{c}{$D$} & \multicolumn{1}{c}{\emph{CC}} & \multicolumn{1}{c}{\emph{d}} \\ \midrule
$\iAPS$ & 180 & 852K  & 305 & 9 & 0.5 & 0.33 & 21  \\
$\iPUBMED$ & 443 & 8M    & 4869    & 37 & 0.8 & 0.36  & 20  \\
$\iDBLP$   & 945 & 4M    & 1413    & 8 & 0.1 & 0.16 & 24  \\
$\iDARTMOUTH$ & 1   & 25K   & 236 & 45    & 410 & 0.51 & 6   \\
$\iUSC$    & 3   & 160K  & 652 & 128   & 510 & 0.49 & 6   \\ 
$\iENRON$ & 87  & 321K  & 1566    & 7 & 0.4 & 0.07 & 20  \\\bottomrule
\end{tabular}
\par
\begin{tablenotes}
\item $|V|$: nr. of nodes; $|E|$: nr. of edges; $\Delta$: max. degree; $\alpha$: avg. degree; $D$: density ($.{10^{-4}}$); $CC$: cluster coeficient; $d$: diameter. The minimum degree in all networks is 1.
\end{tablenotes}
}
\end{threeparttable}
\end{table}
\else 
\begin{table}
        \sisetup{round-mode=places
        ,round-precision=0
        ,scientific-notation=true
    }
\centering
\begin{threeparttable}
    \caption{Characterization of the networks.}
    {
    \label{table:networks_statistics}
    \small
\begin{tabular}{@{}lcccccc@{}}
\toprule
Network  & $\iAPS$ & $\iPUBMED$ & $\iDBLP$ & $\iDARTMOUTH$ & $\iUSC$ & $\iENRON$ \\ \midrule
Nr. of nodes   & 180K     & 443K        & 945K      & 1K             & 3K       & 87K        \\
Nr. of edges      & 852K    & 8M         & 4M       & 25K           & 160K    & 321K      \\
Max. degree & 305     & 4869       & 1413     & 236           & 652     & 1566      \\ \bottomrule
\vspace{-6mm}
\end{tabular}
\par
}
\end{threeparttable}

\end{table}
\fi

Based on the state-of-the-art, we selected seven community detection algorithms for our experiments (see Table~\ref{table:communitydetectionalgorithms}). Our goal is to evaluate the performance of these algorithms before and after our proposed filtering method is applied. 
Due to lack of space, we do not describe these algorithms here, but a detailed discussion of them can be found in the references listed in Table~\ref{table:communitydetectionalgorithms}.

The evaluation of the community detection algorithms was performed as follows. Initially, from the sequence of interactions among the entities, we constructed an aggregated static graph, i.e., a graph that includes all temporal interactions. Then, we executed each algorithm on this graph and evaluated the quality of the detected communities using three different evaluation strategies. 
After that, we used our proposed filtering strategy to remove random interactions from the sequence of interactions and constructed another aggregated static graph from them. 
Finally, we executed and evaluated again all algorithms over this new filtered graph. 

\begin{table}
\centering
\caption{Community detection algorithms.}
\label{table:communitydetectionalgorithms}
\begin{threeparttable}
{
\small
\begin{tabular}{@{}lllcl@{}}
\toprule
\multicolumn{1}{c}{Algorithms} & \multicolumn{1}{c}{M} & Complexity & \multicolumn{1}{c}{Ref.} \\ \midrule
Label Propagation (LP) & N & $O(V)$ & \cite{Raghavan2007} \\
Louvain Modularity (LM) & D & $O(V log V)$ & \cite{blondel20081742-5468-2008-10-P10008} \\
Infomap (IM) & N & $O(V log V)$ & \cite{10.1371/journal.pone.0018209} \\
\makecell{Greedy Optimization of Modularity} (GOM) & D & $O(V log^2(V))$ & \citep{Clauset2004PhysRevE.70.066111} \\
Leading Eigenvector (LE) & D & $O(V^2 log V)$ & \cite{PhysRevE.74.036104} \\
Walktrap (WT) & N & $O(V^2logV)$ & \cite{Pons2005} \\ 
Edge Betweenness (EB) & D & $O(V^3)$ & \cite{NewmanGirvanPhysRevE.69.026113} \\
\bottomrule
\end{tabular}
\begin{tablenotes}
\item M: Algorithm state model (D-deterministic, N-non-deterministic).
\end{tablenotes}
}
\end{threeparttable}
\end{table}

In addition, we analyzed the results of the communities obtained from a graph constructed by a null filtering model, which randomly removes edges from the original network until it reaches the size of the filtered network. 
Given a complete graph $G$ with $n$ nodes and $m$ edges, and the number $k$ of edges to be removed in the stochastic filtering, each edge $m_i$ is removed from the graph with the probability $p$ until to achieve $k$ edges. At the end, all disconnected nodes are removed from $G$. The resulting graph $G_a$ is then used to compare the results achieved by our proposed method with those of the stochastic edge removal method.

Regarding the evaluation strategies, 
modularity is the most used metric to evaluate community detection algorithms \cite{Fortunato201075}. 
To better understand the structure of the communities found by the algorithms, we also use the conductance  metric\footnote{Conductance measures the quality of the cut between a community (a set of nodes) and the rest of the network based on the number of edges outside of that community (inter-cluster conductance) divided by the number of edges inside of that community (intra-cluster conductance) \cite{zaki2014dataminingbook}}, which is a widely-adopted notion of community quality 
\cite{zaki2014dataminingbook}.
Another approach to evaluate improvements in the quality of the detected communities is {to compare them with existing network ground truths~\cite{Peele1602548}.} 
According to Yang and Leskovec \cite{YangLeskovec2015}, a ground truth is built upon particular features of the application domain (e.g., department affiliation), making it possible to divide its entities into groups. However, obtaining a ground truth to evaluate the quality of  clustering algorithms is not an easy task, because it is necessary to have access to data  describing the 
topology of the communities in order to make sense to compare them with the results generated by community detection algorithms. 
We were able to build a ground truth only for the APS dataset, since it was possible to explicitly assign a specific research area to each journal. 
Note that DBLP and PubMed do not provide any explicit information on the research area of the publication venues. 
For the other datasets, 
it was not possible to build a ground truth 
to be compared to the results of the algorithms.

Thus, given a graph \emph{G}, a \emph{ground truth clustering} $P(G)$ and a set of identified communities $C(G)$, {similarity metrics applied to communities are able to} estimate the similarity between $C(G)$ and $P(G)$.
Several metrics are commonly used to measure the similarity between $C(G)$ and $P(G)$~\cite{zaki2014dataminingbook}. Here, due to lack of space, we report results based only on the \textit{split join distance} metric,
which is given by the sum of the projection distance between partitions A and B, being defined, according to Dongen~\citep{Dongen2000}, as: 
\begin{equation}
\rho_A(B)=\sum\limits_{a \in A}{max_{b \in B}|a \cap b |}
\end{equation}
where $|a \cap b|$ denotes the number of common members (overlap) between any subset $a \in A$ and $b \in B$. 

\section{Mining Social Relationships}
\label{sec_methodology_rogue}

In this section, we describe the process proposed for mining social relationships from existing  networks. The main idea behind this process is to remove from the networks noise caused by the presence of random interactions. Thus, our main aim is to reduce errors when associating nodes to communities. 
More formally, we consider a scenario composed of a set of $n$ entities $V=\{v_1, v_2, …, v_n\}$ and an ordered sequence of $m$ interactions $E=\{e_1, e_2, … e_m\}$ (e.g., email exchanges) among the entities of $V$. The $k^{th}$ interaction in $E$ is a tuple $e_k = (t, v_i, v_j)$, where $t$ is the time the interaction occurred, and $v_i$ and $v_j$ are the entities that interacted with each other. The usual approach to detect communities in such a scenario is to construct an aggregated graph from the interactions in $E$ and use this graph as input to any community detection algorithm. Our hypothesis is that if we are able to identify interactions in $E$ that were generated by chance (or are very unlikely to occur in the future), we can remove 
them from $E$ before constructing the aggregated graph, thus improving the quality of the detected communities.

The {main steps 
of our social relationship mining process} 
are: (\textit{i})~characterization of the strength of the relationships from the stream of interactions, (\textit{ii}) removal of the random relationships and (\textit{iii}) reconstruction of the static graph to be used as input to a community detection algorithm. Notice that some techniques used for classifying relationships
from a sequence of interactions may produce different results when the filtered sequence of interactions is used as input for a second time. Because of that, after step (\textit{iii}), step (\textit{i}) may be performed again using the filtered graph as input. This cycle stops when step (\textit{ii}) does not remove any more relationship. Thus, when there are no more random relationships, we obtain a static graph that is composed of only social relationships. Our other hypothesis is that such a static graph allows for more representative communities to be detected by community detection algorithms.

For the purpose of this work, we use the RECAST {classifier}~\cite{VazdeMelo2015} to identify \textit{random} and \textit{social} interactions. \RECAST~classifies relationships (edges in a graph) by assigning a label to each pair of vertices in the graph. Topological and temporal aspects are considered to measure the strength of the edges and then determine which of the following labels will be assigned to each relation: $friend$, $bridge$, $acquaintant$ and $random$. 
For instance, a relationship receives a label $friend$ if the pair of individuals in such a relation has many common friends and regularly interacts over time. {\RECAST~is based on sociological studies that revealed that the topology in which pairs $(v_i, v_j)$ of individuals are involved suggests the strength of the relation between them. 
In this work, this kind of strength is calculated by the neighborhood overlap metric (NO), also known as the Jaccard Index.

\vspace{-1mm}
\section{Experimental Results}\label{sec:results}

This section analyzes the communities obtained by applying the selected algorithms to the six social networks considered. Note that we show the results for the original social networks and also for the filtered ones, i.e., those generated after the complete removal of random relationships. 
By analyzing each social network separately in Fig.~\ref{graphic:ROGUE_Iterations}, it is possible to distinguish them by the number of random relationships identified in each one.
The communication and mobility networks are those with the highest proportion of random relationships. 
As a consequence, nodes that have all their relationships classified as random are disconnected from the network because they do not correspond to members of any community. This means that the individuals represented by these 
nodes do not socially participate in any community. 

\begin{figure}[!t]
     \begin{center}
     \small       \label{graphic:Interactions_by_Class}\includegraphics[width=0.48\textwidth]{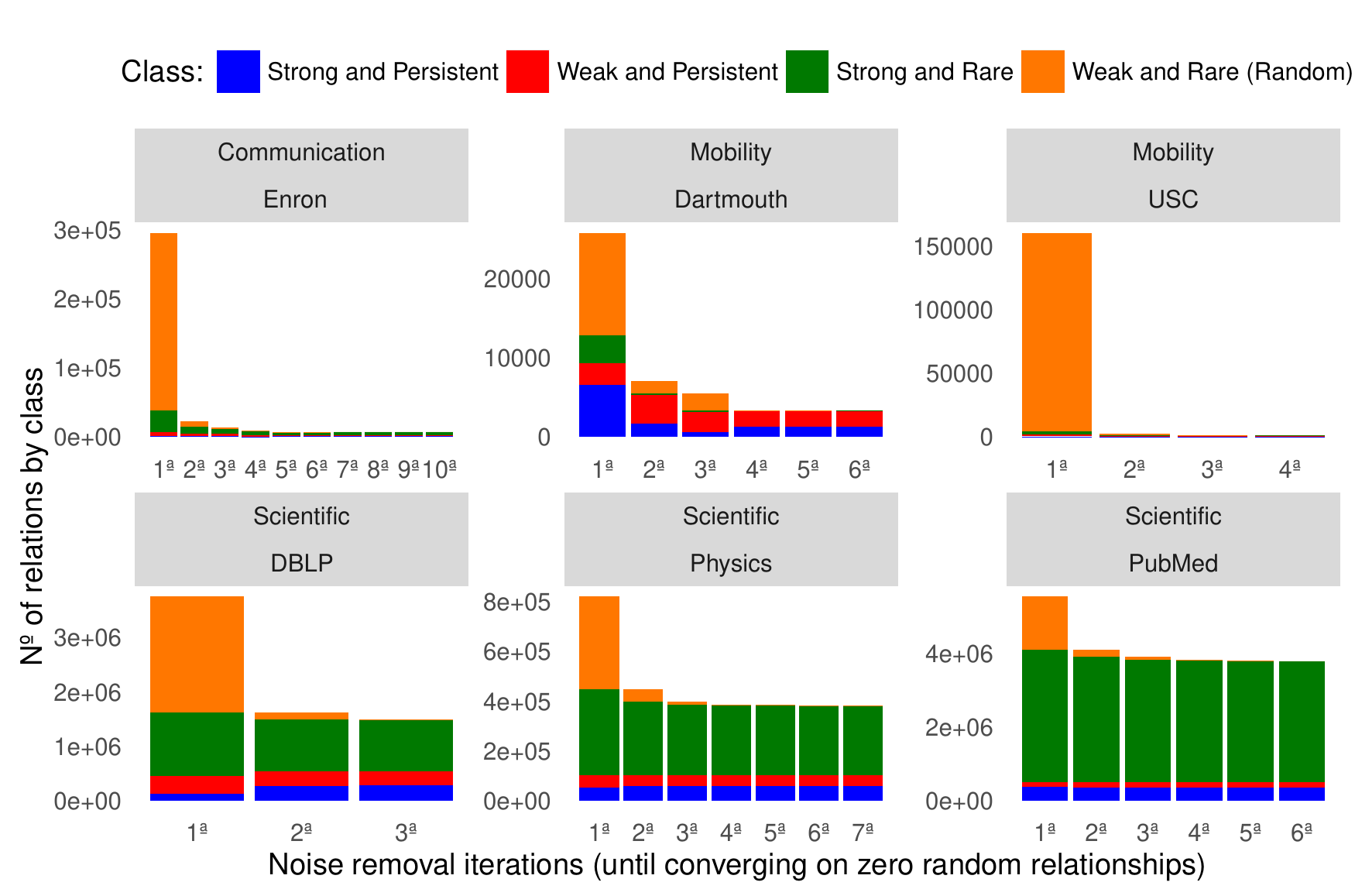}
    \end{center}
    \caption{Relationship classes at the end of each convergence iteration of the social relationship mining process.}
     \label{graphic:ROGUE_Iterations}
     \vspace{-2mm}
\end{figure}

\begin{figure}[!t]
     \begin{center}
            \center
            \includegraphics[width=0.48\textwidth] {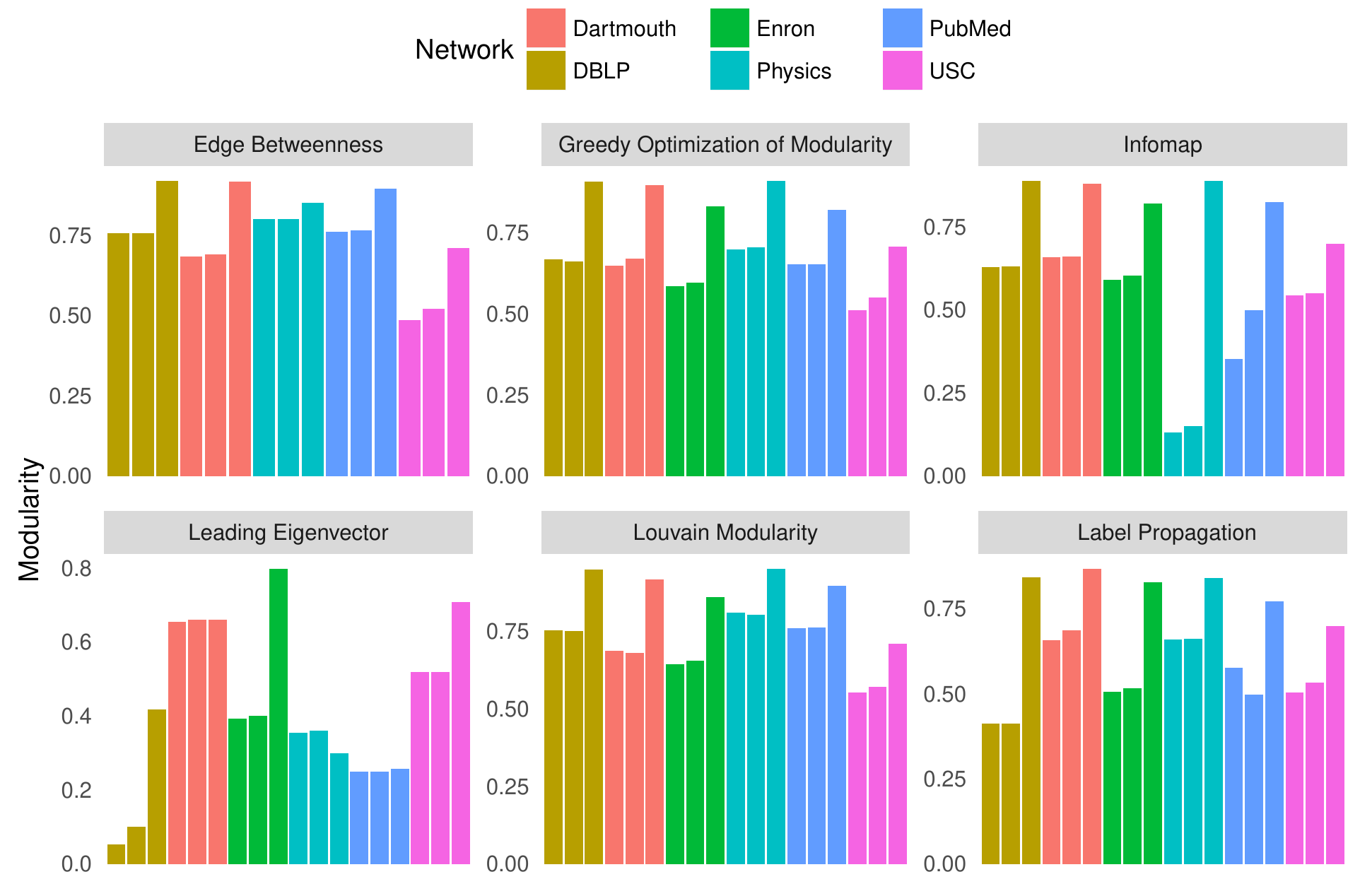}
    \end{center}
    \caption{%
        Modularity of the communities in the original network (left bar), in the network generated by the stochastic process of edge removal {(null model)} (middle bar) and in the network filtered until the last iteration of the \ROGUEStatement~(right bar).}%
     \label{graphic:ROGUE_Modularity_Estimated}
     \vspace{-4mm}
\end{figure}

After the removal of random relationships at each iteration of the~\ROGUEStatement, there was a significant increase in the modularity of the communities detected by the algorithms in each network, as shown in Fig. \ref{graphic:ROGUE_Modularity_Estimated} (right bars). Furthermore, it should be noticed that the structure of the random-edge induced subgraph $R$ (formed only by random edges) of the network $G_c$ (middle bars) is weakly modular when compared to the structure of the original network (left bars). Similar results were obtained for the conductance metric (results omitted).

\begin{figure}[!t]
     \begin{center}
            \center
            \includegraphics[width=0.40\textwidth]{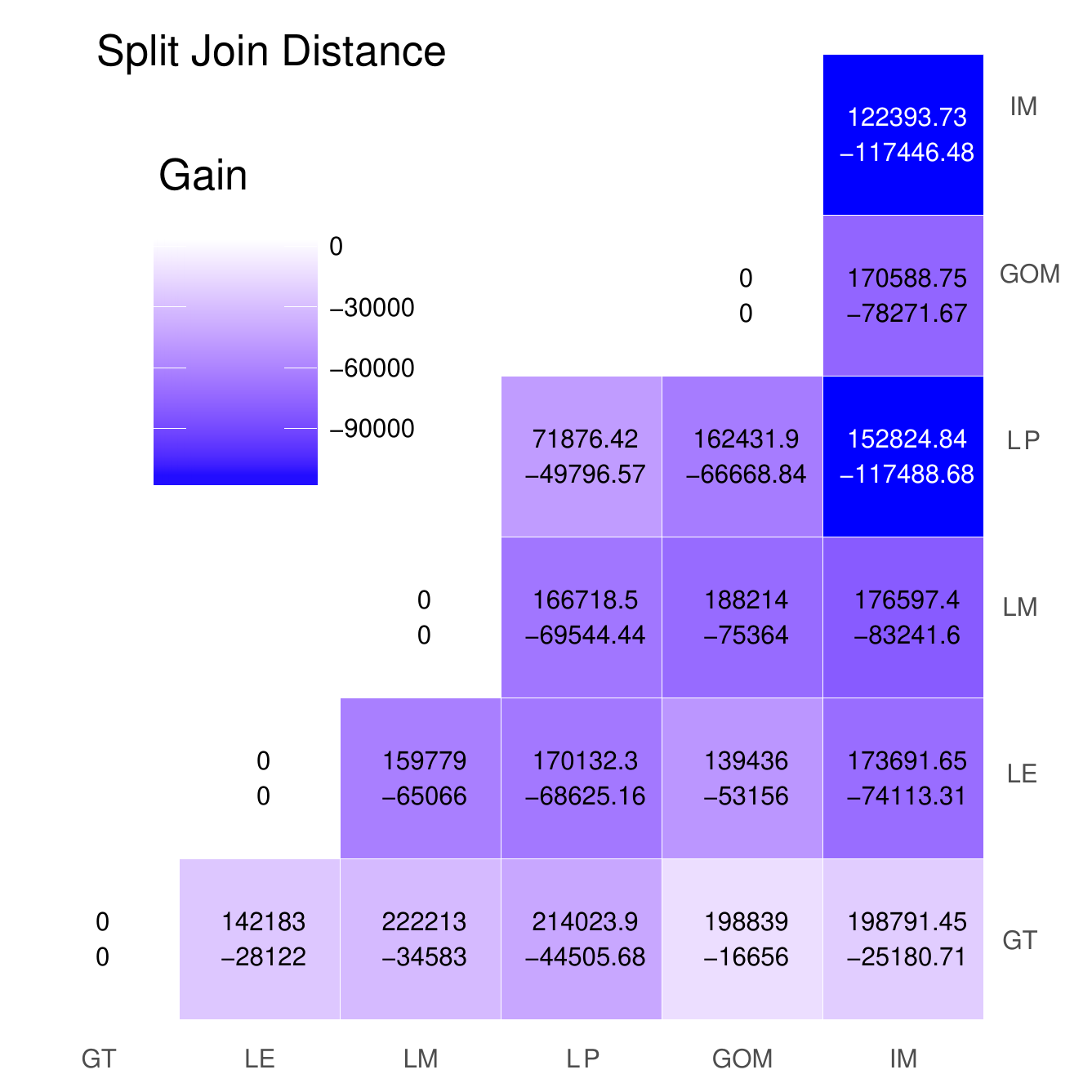}
    \end{center}
    \caption{Split Join Distance (upper value) between each detected community and its ground truth. The gain  after the application of the \ROGUEStatement~is shown below of each similarity value.} 
     \label{figure:aps_compare_gain}
     \vspace{-0.4cm}
\end{figure}

Besides providing a gain of similarity between the network and the ground truth, our \ROGUEStatement~allows the convergence of the community structure, as presented in Fig. \ref{figure:aps_compare_gain}. 
This convergence indicates how much randomness affects a consensus among the diversity of community structure definitions. By removing the noise, we were able to increase the consensus among the algorithms on which communities should be detected. Likewise, such communities have become more similar to the ground truth. 

Finally, the results obtained for the modularity and conductance  metrics in all datasets reveal considerable improvements for the detected communities when random relationships were removed. Regarding the lack of similarity between the detected communities and the APS ground truth, we notice that this was already expected due to the large difference between such communities and those derived from specific metadata (e.g., the researchers' areas of interest), as explained by Hric et al.~\cite{Hric2014}.

\section{Conclusions}\label{sec:conclusion}
The main contribution of this paper is a method for removing noise from temporal social networks that is based on the classification of random relationships and the construction of a static graph composed only of social relationships. 
To evaluate this method, we applied it to six real temporal networks from three distinct domains (scientific collaborations, campus mobility and email communications), and then assessed the quality of their resulting structures.

The application of our method converged by removing noise from all six networks considered in our experiments. 
Furthermore, our results revealed improvements in the communities detected by the state-of-the-art algorithms when compared with those obtained without using our noisy removal method. These improvements correspond to communities with a structure more similar to real ones (ground truths). 
Although our method was applied only to temporal networks and makes use of topological properties to classify relationships, 
it can also be used with other kinds of predictive variables in this task. 

\section*{Acknowledments}
Work supported by project MASWeb (FAPEMIG/PRONEX grant APQ-01400-14) and by the authors' individual grants from CNPq, FAPEMIG and IFNMG.

\bibliographystyle{ACM-Reference-Format}
\bibliography{bibliography}

\end{document}